\documentclass[preprint]{aastex}

\slugcomment{Submitted to ApJ}
\shorttitle{Errata: $EM$s of YSOs in the $\rho$ Oph and Mon R2}
\shortauthors{Imanishi, Kohno, \& Koyama}

\begin{document}

\title{Errata: The X-Ray Emission Measure ($EM$) of the Young Stellar
Objects (YSOs) in the $\rho$ Ophiuchi ($\rho$ Oph) and Monoceros R2 Dark
Clouds (Mon R2)}

\author{Kensuke~Imanishi\altaffilmark{1},
Makoto~Kohno\altaffilmark{1},
and
Katsuji~Koyama\altaffilmark{1}}

\altaffiltext{1}{Department of Physics, Graduate School of Science,
Kyoto University, Sakyo-ku, Kyoto, 606-8502, Japan}
\email{kensuke@cr.scphys.kyoto-u.ac.jp,
kohno@cr.scphys.kyoto-u.ac.jp, koyama@cr.scphys.kyoto-u.ac.jp}

Errors were made in the calculations of $EM$s using the best fit
normalizations in XSPEC. Lists of the errors and corrections are as
follows.

\begin{enumerate}
 \item Tables 1, 3, and 4 in ApJ, 557, 747 and astro-ph/0104190 (YSOs
       in $\rho$ Oph) \\
       All the $EM$s in these tables should be multiplied by $\pi$; the
       log($EM$) in the first raw in Table 1, for example, is
       52.6(52.2--53.2), not 52.1(51.7--52.7).
 \item Figures 7 and 9 in ApJ, 557, 747 and astro-ph/0104190 (YSOs in
       $\rho$ Oph) \\
       The $EM$ plots should be systematically up-shifted by the factor of
       $\pi$.
 \item Table 3 in ApJ, 563, 361 and astro-ph/0108078 (young brown dwarfs
       in $\rho$ Oph)\\
       All the $EM$s in this table should be multiplied by $\pi$; the
       log($EM$) in the first raw, for example, is 52.1(52.0--52.6), not
       51.6(51.5--52.1).
 \item Table 1 in  ApJ, 567, 423  and astro-ph/0110462 (YSOs in Mon R2)  \\
       All the $EM$s in this table should be divided by 10; the log($EM$) in
       the first raw, for example, is 52.5(52.1--52.9), not
       53.5(53.1--53.9).
\end{enumerate}
All the spectral parameters except $EM$ are unchanged.
These errors of $EM$ do not affect any of the conclusions of the
relevant papers.

In addition, Gregorio-Hetem, J., Montmerle, T., Casanova, S., and
Feigelson, E. D.1998, A\&A, 331, 193 should be included in the reference
list of ApJ, 567, 423 (astro-ph/0110462).

\end{document}